\documentclass[
superscriptaddress,
 amsmath,amssymb,
 aps,
]{revtex4-2}

\usepackage{units}
\usepackage{graphicx}
\expandafter\let\csname equation*\endcsname\relax
\expandafter\let\csname endequation*\endcsname\relax

\newcommand{\comment}[1]{}

\begin{document}

\title{Efficient nonlinear compression of a thin-disk oscillator to 8.5\,fs at 55\,W average power}

\author{Gaia Barbiero}
\affiliation{Physics Department, Ludwig-Maximilians-Universit{\"a}t Munich, D-85748 Garching, Germany}
\affiliation{Max Planck Institute of Quantum Optics, D-85748 Garching, Germany}
\author{Haochuan Wang}
\affiliation{Physics Department, Ludwig-Maximilians-Universit{\"a}t Munich, D-85748 Garching, Germany}
\affiliation{Max Planck Institute of Quantum Optics, D-85748 Garching, Germany}
\author{Martin Gra{\ss}l}
\affiliation{Physics Department, Ludwig-Maximilians-Universit{\"a}t Munich, D-85748 Garching, Germany}
\author{Sebastian Gr\"{o}bmeyer}
\affiliation{Physics Department, Ludwig-Maximilians-Universit{\"a}t Munich, D-85748 Garching, Germany}
\author{D\v{z}iugas Kimbaras}
\affiliation{Physics Department, Ludwig-Maximilians-Universit{\"a}t Munich, D-85748 Garching, Germany}
\author{Marcel Neuhaus}
\affiliation{Physics Department, Ludwig-Maximilians-Universit{\"a}t Munich, D-85748 Garching, Germany}
\affiliation{Max Planck Institute of Quantum Optics, D-85748 Garching, Germany}
\author{Vladimir Pervak}
\affiliation{Physics Department, Ludwig-Maximilians-Universit{\"a}t Munich, D-85748 Garching, Germany}
\author{Thomas Nubbemeyer}
\affiliation{Physics Department, Ludwig-Maximilians-Universit{\"a}t Munich, D-85748 Garching, Germany}
\affiliation{Max Planck Institute of Quantum Optics, D-85748 Garching, Germany}
\author{Hanieh Fattahi}
\affiliation{Max Planck Institute for the Science of Light, Staudtstra{\ss}e 2, D-91058 Erlangen, Germany}
\author{Matthias F. Kling}
\email{matthias.kling@lmu.de}
\affiliation{Physics Department, Ludwig-Maximilians-Universit{\"a}t Munich, D-85748 Garching, Germany}
\affiliation{Max Planck Institute of Quantum Optics, D-85748 Garching, Germany}
\affiliation{Center for Advanced Laser Applications, Am Coulombwall 1b, D-85748 Garching, Germany}

\begin{abstract}
We demonstrate an efficient hybrid-scheme for nonlinear pulse compression of high-power thin-disk oscillator pulses to the sub-10\,fs regime. The output of a home-built, 16\,MHz, 84\,W, 220\,fs Yb:YAG thin-disk oscillator at 1030\,nm is first compressed to 17\,fs in two nonlinear multipass cells. In a third stage, based on multiple thin sapphire plates, further compression to 8.5\,fs with 55\,W output power and an overall optical efficiency of 65\% is achieved. By sending the 2.5-cycle pulses into a lithium iodate crystal, we were able to generate ultra-broadband mid-infrared pulses covering the spectral range 2.4-8\,$\mu$m.
\end{abstract}

\maketitle


\section{Introduction}
Nowadays, few-cycle pulses are routinely generated from state-of-the-art high-power, high-energy ultrafast laser amplifier systems, by using parametric amplification\,\cite{budriunas201753,mecseki2019high}, combined with nonlinear compression in noble-gas-filled hollow capillaries\,\cite{nagy2019generation,jeong2018direct,fan202170}, multipass cells\,\cite{balla2020postcompression,muller2021multipass}, or multiplate arrangements\,\cite{lu2014generation,lu2019greater,Seo:2020}. However, such amplifier-based few-cycle systems are usually complex, costly and are typically limited to kHz repetition rates.
While high-power multi-MHz repetition rate-systems have great potential to enhance (time-resolved) spectroscopic applications,
efficient compression schemes to reach below 10 fs are still to be developed. Promising applications for such systems are single-pass\,\cite{hadrich2015exploring} or resonator-enhanced \cite{pupeza2013compact} extreme-ultraviolet generation \cite{gohle2005frequency}, terahertz \cite{meyer2019milliwatt,barbiero2020broadband} and mid-infrared\,(MIR) generation\,\cite{pupeza2020field}. Kerr-lens mode-locked (KLM) Yb:YAG thin-disk (TD) oscillators with their ability in delivering more than 100\,W average power with pulse durations below 200\,fs, combine high peak and high average powers in a relative compact setup \cite{brons2018chapter}. The narrow bandwidth of the gain medium, however, limits the direct use of these powerful sources in ultrafast applications\,\cite{korner2014spectroscopic}, and efficient extra-cavity broadening remains an essential ingredient towards applications that demand ultrashort pulses.

Various approaches utilizing self-phase modulation have been employed so far to overcome the limited bandwidth of KLM Yb:YAG TD oscillators: gas-filled Kagomè photonic crystal fibers (PCF) offer an efficient way for temporal pulse compression, and were shown to yield 9.1\,fs with 14.5\,W average power, in two subsequent stages \cite{mak2015compressing}. However, this approach showed increased susceptibility to damage at high-average and peak powers.
Additionally, fibers are alignment sensitive and couple pointing-drifts to laser power fluctuations. Broadening in multiple plates of bulk material \cite{seidel2016all, seidel2017efficient}, proved to be a valuable alternative reaching 17\,fs. Here, the nonlinear distortion of the beam caused by self-focusing led to a degradation of the beam quality after two stages of bulk media and required additional spatial filtering, limiting the achievable throughput efficiency\,\cite{seidel2016all}. In recent years, efficient and robust compression of KLM Yb:YAG TD oscillators with multipass cells \cite{meyer2019milliwatt, weitenberg2017multi, schulte2016nonlinear, Grobmeyer:2020,  fritsch2018all} has been shown. Multipass cells consist of focusing elements (concave dispersive mirrors) and a nonlinear medium into which the beam is imaged several times. However, the damage threshold of the optical elements and the requirement of accurate dispersion compensation over a large spectral bandwidth limited this approach to the generation of 15\,fs pulses \cite{fritsch2018all}. For further compression towards shorter pulse durations, a different broadening and compression scheme is required.

We implement a hybrid compression scheme for KLM Yb:YAG TD oscillators combining two different approaches in sequential stages: two multipass cells followed by a multiplate compression stage. A similar approach with a multipass cell and multiplate arrangement has been used to compress a mode-locked TD oscillator to 27\,fs at 98\,W\,\cite{tsai2019efficient}.  Here, we demonstrate compression of 220\,fs pulses with 84\,W from a KLM Yb:YAG TD oscillator at 1030\,nm to the sub-10\,fs regime with high optical efficiency of 65\%. One typical application of such a few-cycle near-infrared driving source with high peak power is the generation of broadband carrier-envelope phase (CEP) stable MIR pulses by intra-pulse difference frequency generation (IDFG) \cite{pupeza2015high,zhang2019intra}. Since the shortest wavelength in the generated IDFG output is achieved by mixing the outer wings of the driving laser spectrum, using ultrashort driving pulses is possible to generate MIR radiation in the short-wavelength region. As a proof of principle study, the compressed high-power few-cycle pulses demonstrated here are employed for IDFG in LiIO$_3$, resulting in a broadband CEP-stable MIR continuum spanning the 2.4-8\,$\mu$m spectral range.

\section{Experimental setup and results}

\begin{figure}[t]
\centering
\includegraphics[clip,trim=0.35cm 0.35cm 0.35cm 0.25cm,width=0.6\columnwidth]{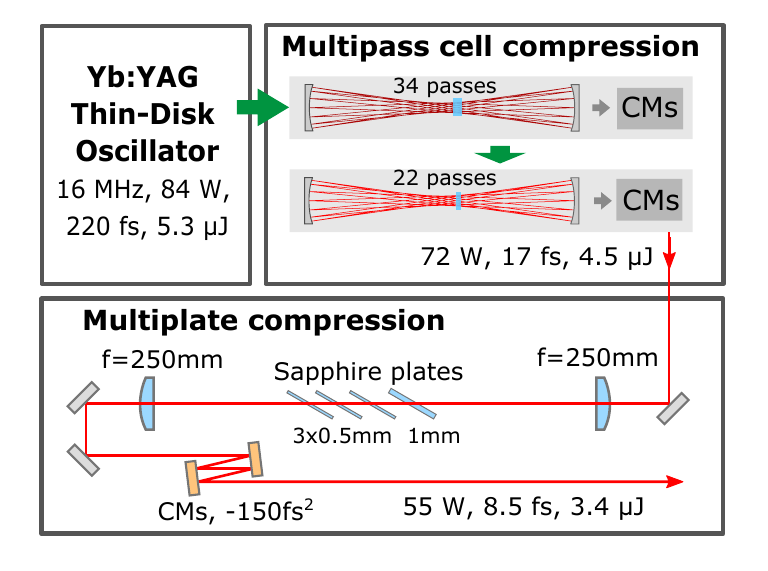}
\caption{Schematic of the hybrid-nonlinear compression setup. The oscillator output is spectrally broadened and compressed in three nonlinear stages. Two consecutive multipass cells compress the oscillator’s output to 17\,fs. The third stage is based on multiplate arrangements, reaching a pulse duration of 8.5\,fs. CMs: chirped mirrors.}
\label{fig:TPs_scheme}
\end{figure}

The layout of the experimental setup is shown in Fig.\ref{fig:TPs_scheme}. The driving laser is a high-power KLM Yb:YAG TD oscillator, delivering 100\,W average power at 16\,MHz repetition rate with a pulse duration of 220\,fs \cite{brons2016powerful}. While 16\,W of the output power is already used for other experiments, the residual 84\,W of the oscillator output are compressed to 17\,fs in two consecutive Herriott-type multipass cells, similar to those depicted in reference\,\cite{barbiero2020broadband}. The first multipass cell is based on two curved dispersive mirrors with radius of curvature (ROC) of 300\,mm, and group delay dispersion (GDD) of -110\,fs$^2$ for 200\,nm spectral bandwidth centered at 1030\,nm. A 6.35-mm thick, anti-reflection (AR)-coated fused silica (FS) plate is used as the nonlinear medium. After 34 passes through the medium, the beam is coupled out and compressed to 49\,fs by 8 bounces on chirped mirrors with a total GDD of -2400\,fs$^2$. The second multipass cell is based on two curved dispersive mirrors with ROC of 300\,mm, and GDD of -60\,fs$^2$ for 450\,nm spectral bandwidth centered at 1030\,nm. A 3\,mm-thick, AR-coated FS plate is used as the nonlinear medium. After 22 passes through the medium, the beam is coupled out and compressed to 17\,fs, close to the Fourier-limit (16\,fs), with 2 bounces on chirped mirrors with a total GDD of -220\,fs$^2$.

Our hybrid compression scheme includes an additional stage to reach the sub-10\,fs regime: a quasi-waveguide in thin plates. As shown in Fig.\ref{fig:TPs_scheme}, four uncoated sapphire plates are placed after the focus of a lens with $f$ = 250\,mm. The plate thicknesses are 1\,mm for the first plate and 0.5\,mm for the other three plates, respectively. After the plates, the beam is re-collimated by another lens and sent to a chirped-mirror compressor with a total GDD of -150\,fs$^2$.

As with other types of quasi-waveguides \cite{meyer2019milliwatt, weitenberg2017multi, schulte2016nonlinear, Grobmeyer:2020,  fritsch2018all}, in this frontend self-phase modulation (SPM) is the main effect exploited to achieve spectral broadening, but nonlinear phase shift per plate is limited such that the spatial and temporal quality of the input beam is maintained after the plates \cite{He:2017, lu2014generation}. For the multiplate waveguide described in this work, the reduction of peak intensity due to material dispersion is not compensated via recompression with dispersive optics as for the multipass cells. Instead, the peak intensity required for SPM is maintained by a reduction of the beam size on successive plates, induced by self-focusing in each plate. The simulated caustic of the self-focused beams was modelled analytically with the \emph{ReZonator2} software\,\cite{ReZonator2}, and is shown in Fig. 2(a), demonstrating the beam-size reduction for consecutive plates.

At the front surface of each plate the beam is divergent with the self-focus located after the back surface. The limit of this waveguide setup is reached when the self-focusing-induced reduction of beam size does not suffice anymore to compensate the material-induced disperson and pulse stretching in the time domain. At this point, the peak power falls below the critical power for self-focusing, such that no further self-focusing can occur.

Different parameters have to be taken into account for the design of the quasi-waveguide. (i) The material should offer a high nonlinear refractive index and high damage threshold for maximum broadening. Four different materials have been tested: fused silica, quartz, YAG and sapphire. With sapphire, the broadest spectrum was achieved without damaging the plate. (ii) The thickness of each plate not only determines SPM, but also the induced self-focusing, and thus the beam size for the subsequent plate. (iii) The beam size on the front surface of each plate is determined by its position with respect to the lens and previous plate(s).

\begin{figure}[t]
\centerline{\includegraphics[clip,trim=0.2cm 0.75cm 0.1cm 0.25cm,width=0.6\columnwidth]{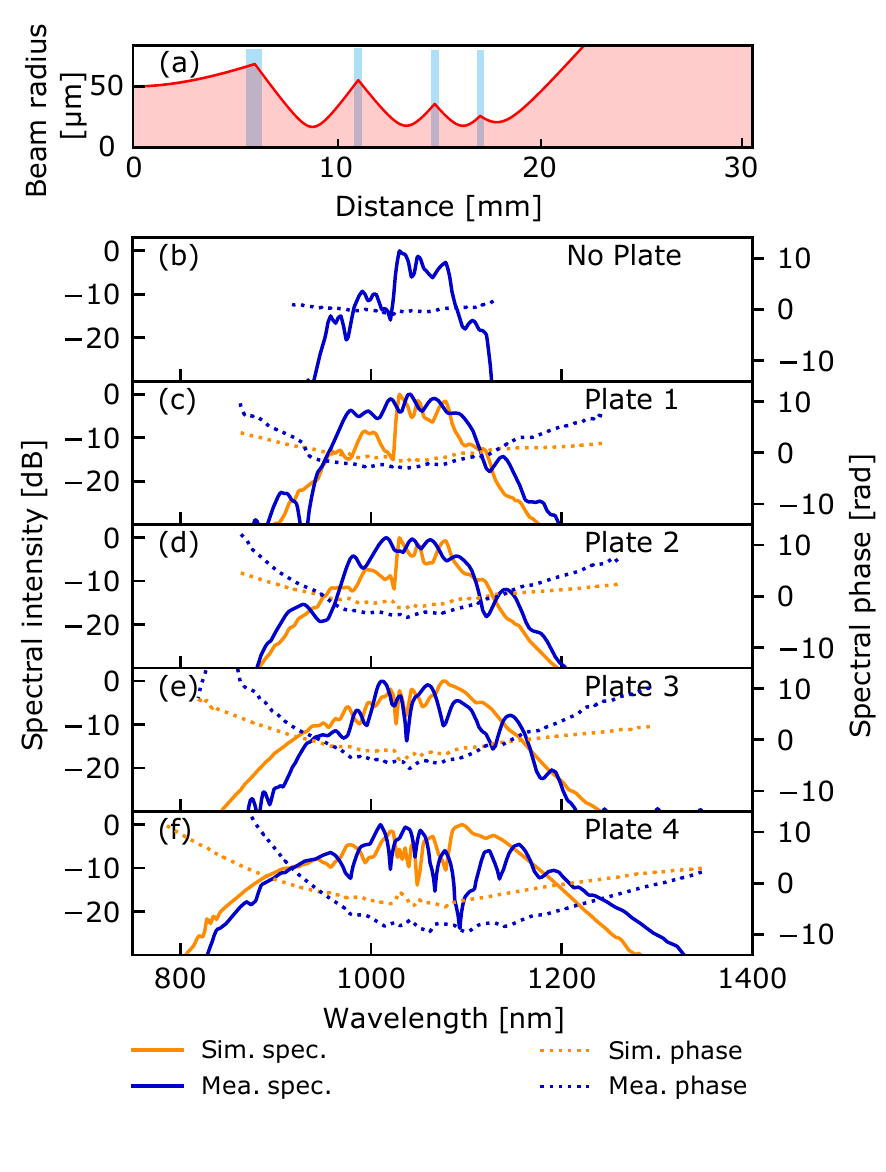}}
\caption{(a)\,Modelled caustic of the input beam along the plates with the \emph{ReZonator2} software \cite{ReZonator2}. FROG retrieved (blue) spectra with spectral phase for the input pulse (b), and after passage through each thin plate (c)-(f). Simulated (orange) spectra with spectral phase employing the Python package PyNLO, are shown for comparison.}\label{fig:TPs_Characterization_0-4}
\end{figure}

\begin{figure}[t]
\centerline{\includegraphics[clip,trim=0 0 0 0,width=0.75\columnwidth]{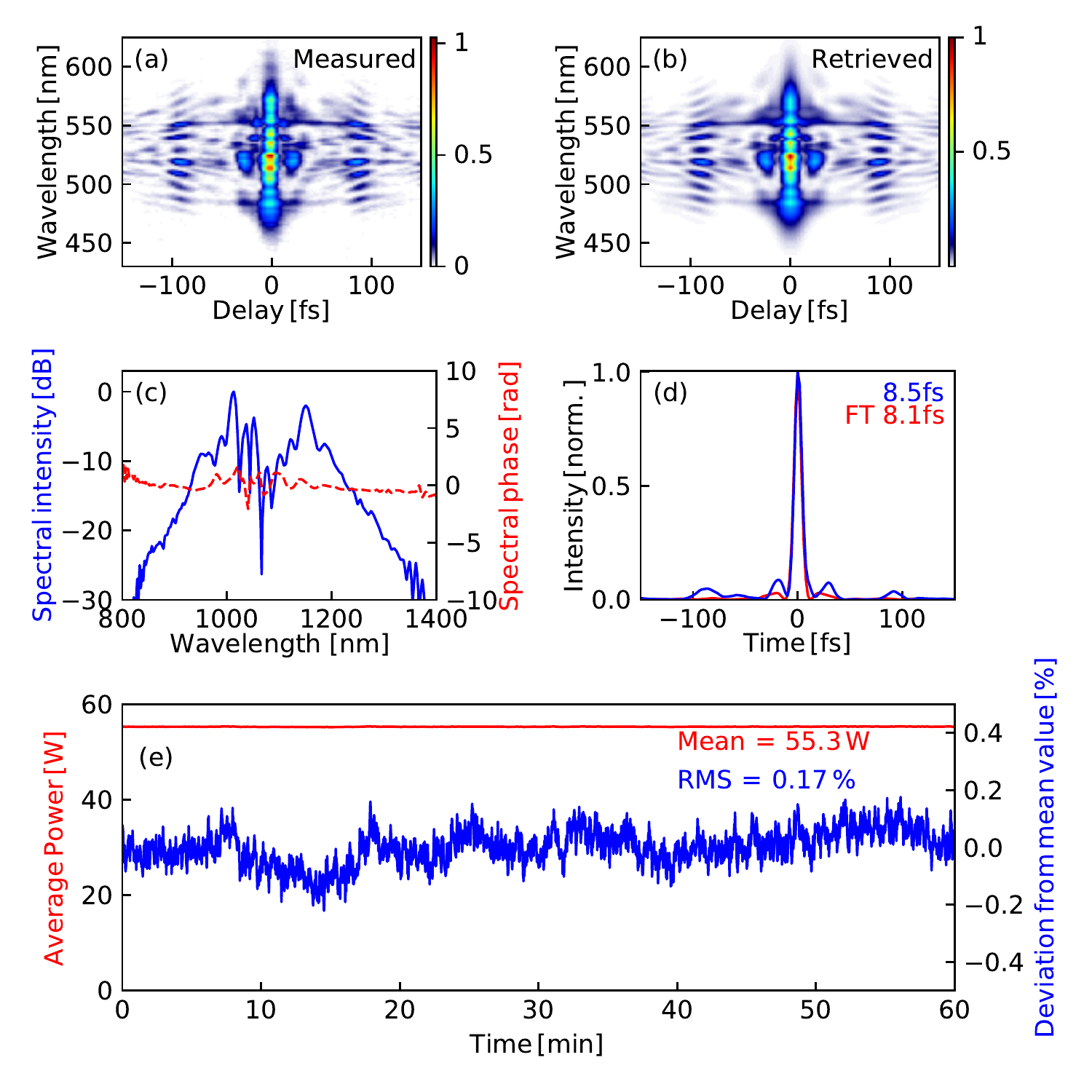}}
\caption{Temporal and power stability characterization of the multiplate setup. (a) Measured and (b) retrieved SH-FROG spectrograms of the thin plates intensity waveguide after the chirp-mirrors compressor (G-error $=$ 0.67\%). (c) Retrieved spectrum. The red dashed line represents the spectral phase. (d) Retrieved temporal profile. (e) Average power stability measured over 1\,h with 1\,Hz sampling frequency.}
\label{fig:TPs_Characterization}
\end{figure}

Different combinations of sapphire plates with thicknesses ranging from 0.25\,mm to 5\,mm have been tested. The highest spectral broadening is achieved with a divergent input beam. The higher the divergence, the higher the required positioning precision for the plate along the beam path. This is a consequence of the smaller range between the two beam radii where the intensity is sufficient for self-focusing and where the intensity is above the damage threshold of the plate. The $\mu$J-level pulse energy after the multipass cells implied a small beam size on the order of 100 microns for sufficient self-focusing effects. For this reason, each plate is placed close to the focus, and the relative distances between the plates are very small, on the order of 5\,mm. This strongly limited the space between the plates, and therefore all plates are positioned at the same Brewster angle and not at complementary Brewster angles as in e.g. \cite{Seo:2020}.

The focal length of the lens was chosen to be 250\,mm and the best broadening result after the first plate was obtained with a 1-mm thick sapphire plate, where damage of the plate only occurs at the focus. The subsequent plates are positioned as follows: each plate is placed after the focus and moved towards the focus till the highest broadening is reached without damaging of the plate or distortion of the output beam. The divergent beam after the last plate is then re-collimated by an identical lens ($f$ = 250\,mm), and a portion of the broadened spectrum is characterized by a spectrometer for comparison. The material dispersion of the plates stretches the pulses to 80\,fs. The output beam is re-compressed by 4 bounces on chirped mirrors with a total GDD of -150\,fs$^2$. The total optical efficiency of the multiplate stage including the chirped mirrors is 80\%.

The effect of each plate with respect to spectral broadening and phase is shown in Fig.\ref{fig:TPs_Characterization_0-4}(b)-(f) (blue). The spectral broadening was simulated by solving the nonlinear Schr{\"o}dinger equation using the fourth-order Runge–Kutta in the Interaction Picture (RK4IP) method using the Python package PyNLO\,\cite{hult2007fourth}, assuming constant beam diameters within the plates. The comparison of simulated\,(orange) and experimental\,(blue) spectra in Fig. \ref{fig:TPs_Characterization_0-4}(b)-(f) shows good agreement up to the third plate. After this point, for accurate comparison between experimental data and simulations, careful full-field 4D simulation taking into account spatial-temporal coupling would be required \cite{Hanna:2017}, which were outside the scope of this work.

The compressed pulses after the chirped mirrors are characterized by second-harmonic frequency resolved optical gating (SH-FROG), employing a 10\,$\mu$m-thick beta-barium borate crystal. The measured and retrieved spectrograms, the spectrum and the temporal profile are shown in Fig. \ref{fig:TPs_Characterization}(a)-(d). The temporal profile shows that the pulses are re-compressed to 8.5\,fs (FWHM), close to the Fourier-limit of 8.1\,fs. In addition to a remarkable temporal compression, high power and beam pointing stability after the multiplate setup are measured. The average power root-mean-square (RMS) fluctuations are only 0.17\% over 1 hour (Fig. \ref{fig:TPs_Characterization}(e)). The RMS beam pointing fluctuations are 0.7\% within 1 hour. The output beam profile has an M$^2$ value of 1.29 and 1.40, in $x$ and $y$ directions, respectively (ISO 11146 measured with M$^2$-200s, Ophir-Spiricon LLC).

\section{Broadband mid-infrared generation}

As a simple approach to generate broadband MIR pulses, IDFG driven by femtosecond pulses \cite{pires2015ultrashort,pupeza2015high} has intrinsic CEP-stability yielding highly reproducible MIR waveforms. Here, we demonstrate that the compressed 8.5-fs pulses from our oscillator system can generate an MIR continuum spanning from 2.4 to 8\,$\mu$m.

\begin{figure}[t]
\centerline{\includegraphics[clip,trim=0.75cm 0.4cm 1cm 0,width=0.6\columnwidth]{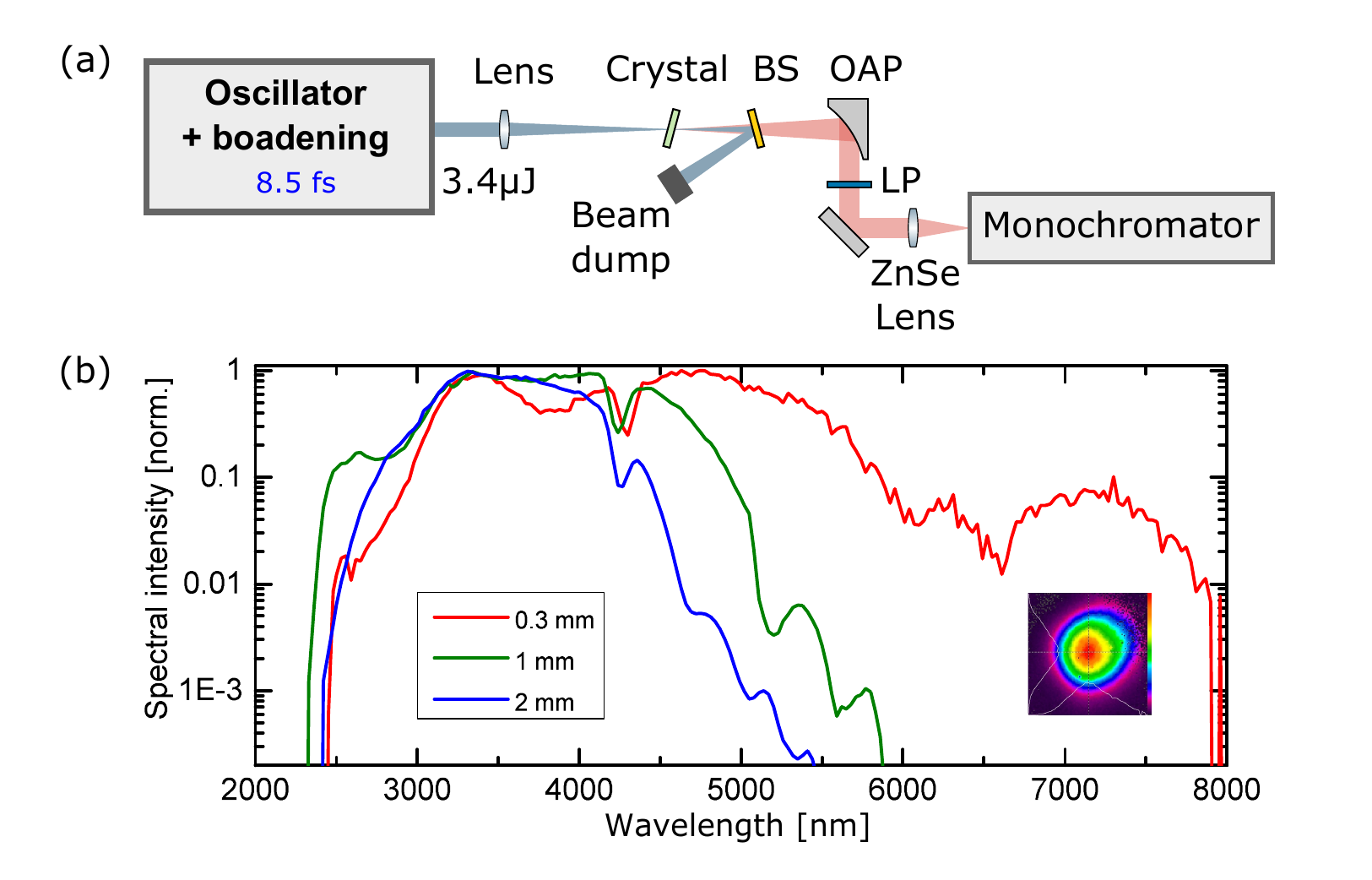}}
\caption{Broadband short-range MIR generation by IDFG. (a) Schematic of MIR generation and characterization setup.  (b) The generated MIR spectrum in 0.3\,mm, 1\,mm, and 2\,mm LiIO$_3$ crystals. Inset: measured beam profile after 2\,mm crystal and a 2400\,nm long-pass filter (LP2400- 33-969, Edmund Optics Ltd.). BS: beam splitter, OAP: off-axis parabola, LP: long-pass filter.}
\label{fig:MIR}
\end{figure}

As depicted in Fig.\ref{fig:MIR}(a), the 8.5\,fs pulses are focused into LiIO$_3$ crystal with an $f$\,=\,300\,mm lens. The crystal is oriented to evenly distribute the p-polarized input pulses along the extraordinary and the ordinary axes for type-I phase matching. The generated MIR radiation is separated by a tailored ZnSe beam splitter (UltraFast Innovations GmbH, F3-S161108) and then re-collimated by a gold-coated off-axis parabolic mirror with focal length of 101.6\,mm. A 2400\,nm long-pass filter is used to remove the residual pump radiation transmitted through the beam splitter. After the long-pass filter the MIR radiation is characterized by a power meter and a monochromator (Newport Cornerstone 260).

The crystals (EKSMA Optics, UAB) used in this work had broadband AR coatings for the incident NIR light and different thicknesses: 2\,mm, 1\,mm, and 0.3\,mm. The highest average powers measured before damage of the crystal, are 72\,mW, 16\,mW, and 1.1\,mW, respectively. The corresponding IDFG spectra are shown in Fig.\ref{fig:MIR}(b). To suppress second-order diffraction of the monochromator, three long-pass filters with cut-on wavelengths at 2.4, 4.5, and 7.3\,$\mu$m are used and the spectra merged \cite{zhang2019intra}. For the 0.3\,mm crystal, the spectral coverage spanned almost two octaves, from 2.4\,$\mu$m to 8\,$\mu$m within a $10^{-3}$ dynamic range. While in previous IDFG experiments driven by KLM Yb:YAG TD oscillators with 16 fs pulses only wavelengths down to 3\,$\mu$m could be covered\,\cite{zhang2019intra}, the current results show spectral coverage down to 2.4\,$\mu$m.

\section{Conclusions}
In conclusion, an efficient hybrid compression scheme for KLM Yb:YAG TD oscillators was demonstrated. The setup consisted of two consecutive multipass cells and a waveguide-type multiplate stage. With an overall efficiency of 65\%, we achieved 8.5\,fs pulses at 16\,MHz with 55\,W average power and excellent long-term power and beam pointing stability from a simple setup. The generated few-cycle pulses from KLM Yb:YAG TD oscillators open up many applications, such as the generation ultrabroadband MIR radiation for time-resolved spectroscopy. Intra-pulse difference frequency generation in LiIO$_3$ generated a broadband MIR continuum spanning from 2.4 to 8\,$\mu$m. The high repetition rate combined with the high intensity and the remarkable stability of this system paves the way for future applications that require high intensities, such as further spectral broadening, broadband frequency up/down-conversion, and new applications in ultrafast spectroscopy.


\section*{Funding}
Deutsche Forschungsgemeinschaft (DFG) (LMUexcellent); Max-Planck Gesellschaft (MPG) (IMPRS-APS, MPSP, Fellow Program).

\section*{Acknowledgements}
We are grateful to Ferenc Krausz for his support and for providing suitable laboratory space. We acknowledge help by Jonathan Brons for his support regarding the oscillator, and Kafai Mak and Alexander Weigel for useful discussions and proofreading. G.B., H.W. and D.K. are grateful for support by the International Max Planck Research School on Advanced Photon Science (IMPRS-APS). D.K. and M.F.K. acknowledge support by the Max Planck School of Photonics (MPSP), and M.F.K. is grateful for support by the Max Planck Society within their Max Planck Fellow program.

\section*{Disclosures} The authors declare no conflicts of interest.

\section*{Data Availability Statement}
The data will be provided upon reasonable request from the authors.

\end{document}